\documentclass[12pt]{article}
\usepackage{epsfig}
\usepackage{amsfonts}
\begin{document}
\begin{titlepage}
\begin{center}

{\Large Superstatistics: Recent developments and applications}

\vspace{2.cm} {\bf Christian Beck}

\hspace{2cm}

School of Mathematical Sciences, Queen Mary, University of
London, Mile End Road, London E1 4NS, UK

\vspace{2cm}

\end{center}

\abstract{We review some  recent developments which
make use of the concept of `superstatistics', an effective
description for nonequilibrium systems with a varying intensive
parameter such as the inverse temperature. We describe how the
asymptotic decay of stationary probability densities can be
determined using a variational principle, and present some
new results on the typical
behaviour of correlation functions in dynamical superstatistical
models. We briefly describe some recent applications of the
superstatistics concept in hydrodynamics, astrophysics,
and finance.}

\end{titlepage}

\section{Introduction}

Complex nonequilibrium systems often exhibit dynamical behaviour
that is characterized by spatio-temporal fluctuations of an
intensive parameter $\beta$. This intensive parameter may be the
inverse temperature, or an effective friction constant, or the
amplitude of Gaussian white noise, or the energy dissipation in
turbulent flows, or simply a local variance parameter extracted
from a signal. A nonhomogeneous spatially extended system with
fluctuations in $\beta$ can be thought of as consisting of a
partition of spatial cells with a given $\beta$ in each cell. If
there is local equilibrium in each cell (so that statistical
mechanics can be applied locally), and if the fluctuations of
$\beta$ evolve on a sufficiently large time scale, then in the
long-term run the entire system is described by a superposition
of different Boltzmann factors with different $\beta$, or in
short, a `superstatistics' \cite{beck-cohen}. The superstatistics
approach has been the subject of various recent papers
\cite{cohen,boltzmann-m,beck-su,touchette,sattin04,supergen,hanggi,abearbi,ryazanov,aringazin,souza,souza2}.
Superstatistical techniques can be successfully applied to a
variety of physical problems, such as
Lagrangian\cite{beck03,reynolds} and Eulerian
turbulence\cite{castaing,beck-physica-d}, 
defect turbulence\cite{daniels}, cosmic ray statistics\cite{cosmic},
plasmas\cite{sattin02}, 
statistics of wind velocity differences\cite{rapisarda,rap2}
and mathematical 
finance\cite{bouchard,ausloos,econo}. Experimentally measured
non-Gaussian stationary distributions with `fat tails' can often
be successfully described by simple models that exhibit a
superstatistical spatio-temporal dynamics.

If the intensive parameter in the various cells is distributed
according to a particular probability distribution, the
$\chi^2$-distribution, then the corresponding superstatistics,
obtained by integrating over all $\beta$, is given by Tsallis
statistics \cite{tsa1,tsa2,tsa3,abe}. For other distributions of
the intensive parameter $\beta$, one ends up with more general
superstatistics, which contain Tsallis statistics as a special
case. Generalized entropies (analogues of the Tsallis entropies)
can also be defined for general 
superstatistics\cite{abearbi,souza,souza2,chavanis} 
and are indeed a useful tool.
The ultimate goal is to proceed from simple models to more general
versions of statistical mechanics, which are applicable to wide
classes of complex nonequilibrium systems, thus further
generalizing Tsallis' original ideas\cite{tsa1}.

This paper is organized as follows:

First, we briefly review the superstatistics concept. We then show
how one can deduce the asymptotic decay rate of the stationary
probability densities of general superstatistics from a
variational principle\cite{touchbeck}. In section 4 we consider
dynamical realization of superstatistics and investigate the
typical behaviour of correlation functions.
Many different types of decays of correlations (e.g.\ power law,
stretched exponentials) are possible.
Section 5 summarizes
some recent applications of the superstatistics
concept in hydrodynamics,
astrophysics, and finance.

\section{What is superstatistics?}

The superstatistics approach is applicable to a large variety of
driven nonequilibrium systems with spatio-temporal fluctuations
of an intensive parameter $\beta$, for example, the inverse
temperature. Locally, i.e.\ in spatial regions (cells) where
$\beta$ is approximately constant, the system is described by
ordinary statistical mechanics, i.e.\ ordinary Boltzmann factors
$e^{-\beta E}$, where $E$ is an effective energy in each cell. In
the long-term run, the system is described by a spatio-temporal
average of various Boltzmann factors with different $\beta$. One
may define an effective Boltzmann factor $B(E)$ as
\begin{equation}
B(E) =\int_0^\infty f(\beta) e^{-\beta E}d\beta  =\langle
e^{-\beta E} \rangle, \label{1}
\end{equation}
where $f(\beta)$ is the probability distribution of $\beta$ in the
various cells. For so-called type-A
superstatistics\cite{beck-cohen}, one normalizes this effective
Boltzmann factor and obtains the stationary long-term probability
distribution
\begin{equation}
p(E)=\frac{1}{Z}B(E),
\end{equation}
where
\begin{equation}
Z=\int_0^\infty B(E)dE.
\end{equation}
For type-B superstatistics, the $\beta$-dependent normalization
constant of each local Boltzmann factor is included into the
averaging process. In this case the invariant long-term
distribution is given by
\begin{equation}
p(E)=\int_0^\infty f(\beta) \frac{1}{Z(\beta)}e^{-\beta E}d\beta
, \label{2}
\end{equation}
where $Z(\beta)$ is the normalization constant of $e^{-\beta E}$
for a given $\beta$. Eq.~(\ref{2}) is just a simple consequence
of calculating marginal distributions. Type-B superstatistics can
easily be mapped into type-A superstatistics by re-defining
$f(\beta)$.

A superstatistics can be dynamically realized by considering
Langevin equations whose parameters fluctuate on a relatively
large time scale (see \cite{prl} for details). For example, for
turbulence applications one may consider a superstatistical
extension of the Sawford model of Lagrangian
turbulence\cite{beck03,reynolds,sawford}. This model consists of
suitable stochastic differential equations for the position,
velocity and acceleration of a Lagrangian test particle in the
turbulent flow, and the parameters of this model then become
random variables as well. Experimental data are well reproduced by
these types of models.

Often, a superstatistics just consists of a superposition
of Gaussian distributions with varying variance. The parameter
$\beta$ can then be estimated from an experimentally measured
signal $u(t)$ as
\begin{equation}
\beta= \frac{1}{\langle u^2 \rangle_T -\langle u \rangle^2_T},
\label{locva}
\end{equation}
where $\langle ...\rangle_T$ denotes an average over a finite time
interval $T$ of the signal, corresponding to the `cell size' of
the superstatistics. It is then easy to make histograms of $\beta$
and thus empirically determine $f(\beta)$.

\section{Asymptotic behaviour for large energies}

Superstatistical invariant densities, as given by eq.(\ref{1}) or
(\ref{2}), typically exhibit `fat tails' for large $E$, but what
is the precise functional form of this large energy behaviour? The
answer depends on the distribution $f(\beta)$ and can be obtained
from a variational principle. Details are described
in \cite{touchbeck}, here we just summarize
some results. For large $E$ we
may use the saddle point approximation and write
\begin{eqnarray}
B(E) &=&\int_0^\infty f(\beta )e^{-\beta E}d\beta  \nonumber \\
&=&\int_0^\infty e^{-\beta E+\ln f(\beta )}d\beta  \nonumber \\
&\sim &e^{\sup_\beta \{-\beta E+\ln f(\beta )\}}  \nonumber \\
&=&e^{-\beta _EE+\ln f(\beta _E)}  \nonumber \\
&=&f(\beta _E)e^{-\beta _EE}, \label{5}
\end{eqnarray}
where
\begin{equation}
\beta _E= \sup_\beta \{-\beta E+\ln f(\beta )\}.  \label{lf3}
\end{equation}
The expression \begin{equation} \sup_\beta \{-\beta E+\ln f(\beta
)\}
\end{equation}
corresponds to a Legendre transform of $\ln f(\beta )$. The
result of this transform is a function of $E$ which can be
thought of as representing a kind of entropy function if we
consider the function $\ln f(\beta )$ to represent a free
energy function. This entropy function, however, is different
from other entropy functions used e.g.\ in nonextensive
statistical mechanics. It describes properties related
to the
fluctuations of inverse temperature.

In the case where $f(\beta )$ is smooth and has only a single
maximum we can obtain the supremum by differentiating, i.e.\
\begin{equation}
\sup_\beta \{-\beta E+\ln f(\beta )\}=-\beta _EE+\ln f(\beta _E)
\end{equation}
where $\beta _E$ satisfies the differential equation
\begin{equation}
0=-E+(\ln f(\beta ))^{\prime }=-E+\frac{f^{\prime }(\beta
)}{f(\beta )}. \label{lf1}
\end{equation}
By taking into account the next-order contributions around the
maximum, eq.~(\ref{5}) can be improved to
\begin{equation}
B(E)\sim \frac{f(\beta _E)e^{-\beta _EE}}{\sqrt{-(\ln f(\beta
_E))^{\prime \prime }}}.
\end{equation}

Let us consider a few examples. Consider an $f(\beta)$ of the
power-law form
$f(\beta )\sim \beta ^\gamma $, $\gamma >0$ for small $\beta$. An
example is a $\chi ^2$ distribution of $n$ degrees of
freedom\cite{prl,wilk},
\begin{equation}
f(\beta )=\frac 1{\Gamma (\frac n2)}\left( \frac n{2\beta
_0}\right) ^{n/2}\beta ^{n/2-1}e^{-\frac{n\beta }{2\beta _0}},
\label{chi2}
\end{equation}
($\beta _0\geq 0$, $n>1$) which behaves for $\beta \to 0$ as
\begin{equation}
f(\beta )\sim \beta ^{n/2-1},
\end{equation}
i.e.\
\begin{equation}
\gamma = \frac{n}{2}-1.
\end{equation}
Other examples exhibiting this power-law form are
$F$-distributions\cite{beck-cohen,sattin02}. With the above
formalism one obtains from eq.~(\ref{lf1})
\begin{equation}
\beta _E =\frac \gamma E
\end{equation}
and
\begin{equation}  B(E) \sim E^{-\gamma -1}.
\end{equation}
These types of $f(\beta)$ form the basis for power-law 
generalized Boltzmann factors ($q$-exponentials) $B(E)$, with the
relation \cite{tsa1,tsa2,tsa3,abe}
\begin{equation}
\gamma +1 = \frac{1}{q-1}. \label{here}
\end{equation}

Another example would be an $f(\beta)$ which for small $\beta$
behaves as $f(\beta )\sim e^{-c/\beta }$, $c>0$. In this case one
obtains
\begin{equation}
\beta_E=\sqrt{\frac{c}{E}}
\end{equation}
and \begin{equation}
 B(E) \sim
E^{-3/4}e^{-2\sqrt{cE}}.
\end{equation}
The above example can be generalized to stretched exponentials:
For $f(\beta)$ of the form $f(\beta )\sim e^{-c\beta ^\delta }$
one obtains after a short calculation
\begin{equation}
\ \beta _E =\left( \frac E{c|\delta |}\right) ^{1/(\delta -1)}
\end{equation}
and
\begin{equation}
B(E) \sim E^{(2-\delta )/(2\delta -2)}e^{aE^{\delta /(\delta -1)}},
\end{equation}
where $a$ is some factor depending on $\delta$ and $c$.

Of course which type of $f(\beta)$ is relevant depends on the
physical system under consideration. For many problems in
hydrodynamic turbulence, log-normal superstatistics seems to be
working as a rather good approximation. In this case $f(\beta)$ is
given by
\begin{equation}
f(\beta)= \frac{1}{\sqrt{2\pi}s\beta}\exp \left\{
-\frac{(\log \beta - m)^2}{2s^2} \right\} ,
\end{equation}
where $s$ and $m$ are parameters
\cite{beck-cohen,beck03,reynolds,castaing,beck-physica-d,jungswinney}.

\section{Superstatistical correlation functions}
To obtain statements on correlation functions, one has to
postulate a concrete dynamics that generates the superstatistical
distributions. The simplest dynamical model of this kind is a
Langevin equation with parameters that vary on a long time scale,
as introduced in \cite{prl}.

Let us consider a Brownian particle of mass $m$ and a Langevin equation of
the form
\begin{equation}
\dot{v}=-\gamma v + \sigma L(t),
\end{equation}
where $v$ denotes the velocity of the particle, and $L(t)$ is
normalized Gaussian white noise with the following expectations:
\begin{eqnarray}
\langle L(t) \rangle &=&0 \\
\langle L(t)L(t') \rangle &=&\delta (t-t').
\end{eqnarray}
We assume that the parameters $\sigma$ and $\gamma$ are constant
for a sufficiently long time scale $T$, and then change to new
values, either by an explicit time dependence, or by a change of
the environment through which the Brownian particle moves. Formal
identification with local equilibrium states in the cells
(ordinary statistical mechanics at temperature $\beta^{-1}$)
yields during the time scale $T$ the relation\cite{vKa}
\begin{equation}
\langle v^2 \rangle =\frac{\sigma^2}{2\gamma}=\frac{1}{\beta m}
\label{einstein}
\end{equation}
or
\begin{equation}
\beta = \frac{2}{m} \frac{\gamma}{\sigma^2}.  \label{beta}
\end{equation}
Again, we emphasize that after the time scale $T$, $\gamma$ and
$\sigma$ will take on new values. During the time interval $T$,
the probability density $P(v,t)$ obeys the Fokker-Planck equation
\begin{equation}
\frac{\partial P}{\partial t}=\gamma \frac{\partial (vP)}{\partial
v}+ \frac{1}{2} \sigma^2 \frac{\partial^2 P}{\partial v^2}
\end{equation}
with the local stationary solution
\begin{equation}
P(v|\beta)=\sqrt{\frac{m\beta}{2\pi}} \exp \left\{-\frac{1}{2}
\beta mv^2 \right\} \label{28}.
\end{equation}
In the adiabatic approximation, valid for large $T$, one asumes
that the local equilibrium state is reached very fast so that
relaxation processes can be neglected. Within a 
cell in local equilibrium the correlation function is given by
\cite{vKa}
\begin{equation}
C(t-t'|\beta)=\langle v(t) v(t')\rangle =
\frac{1}{m\beta}e^{-\gamma |t-t'|}.
\end{equation}
Clearly, for $t=t'$ and setting $m=1$ we have
\begin{equation}
\beta= \frac{1}{\langle v^2 \rangle_T},
\end{equation}
in agreement with eq.~(\ref{locva}).

It is now interesting to see that the long-term invariant
distribution $P(v)$, given by
\begin{equation}
P(v)=\int_0^\infty f(\beta) P(v|\beta) d\beta \label{31}
\end{equation}
depends only on the probability distribution of
$\beta=\frac{2}{m} \frac{\gamma}{\sigma^2}$ and not on that of the
single quantities $\gamma$ and $\sigma^2$. This means, one can
obtain the same stationary distribution from different dynamical
models based on a Langevin equation with fluctuating parameters.
Either $\gamma$ may fluctuate, and $\sigma^2$ is constant, or the
other way round. On the other hand, the superstatistical
correlation function
\begin{equation}
C(t-t')=\int_0^\infty f(\beta) C(t-t'|\beta) d\beta =\frac{1}{m}
\int_0^\infty f(\beta) \beta^{-1}e^{-\gamma |t-t'|}d\beta
\label{32}
\end{equation}
can distinguish between these two cases. The study of correlation
functions thus yields more information for any superstatistical
model.

Let illustrate this with a simple example. Assume that $\sigma$
fluctuates and $\gamma$ is constant such that $\beta=\frac{2}{m}
\frac{\gamma}{\sigma^2}$ is $\chi^2$-distributed. Since $\gamma$
is constant, we can get the exponential $e^{-\gamma |t-t'|}$ out
of the integral in eq.~(\ref{32}), meaning that the
superstatistical correlation function still decays in an
exponential way:
\begin{equation}
C(t-t') \sim e^{-\gamma |t-t'|}. \label{expo}
\end{equation}
On the other hand, if $\sigma$ is constant and $\gamma$ fluctuates
and $\beta$ is still $\chi^2$-distributed with degree $n$, we get
a completely different answer. In this case,
in the adiabatic approximation, the integration
over $\beta$ yields a power-law decay of $C(t-t')$:
\begin{equation}
C(t-t') \sim |t-t'|^{-\eta}, \label{power}
\end{equation}
where
\begin{equation}
\eta =\frac{n}{2}-1 \label{eta}
\end{equation}
Note that this decay rate is different from the asymptotic power law
decay rate of the invariant density $P(v)$, which, using
(\ref{28}) and (\ref{31}), is given by
$P(v)\sim v^{-2/(q-1)}$, with
\begin{equation}
\frac{1}{q-1} =\frac{n}{2}+\frac{1}{2}. \label{qqq}
\end{equation}
In general, we may generate many different types of correlation
functions for general choices of $f(\beta)$. By letting both
$\sigma$ and $\gamma$ fluctuate we can also construct
intermediate cases between the exponential decay (\ref{expo}) and
the power law decay (\ref{power}), so that strictly speaking we
only have the inequality
\begin{equation}
\eta \geq \frac{n}{2}-1,
\end{equation}
depending on the type of parameter fluctuations considered.

One may also proceed to the position
\begin{equation}
x(t)=\int_0^t v(t')dt'
\end{equation}
of the test particle. One has
\begin{equation}
\langle x^2(t) \rangle = \int_0^t \int_0^t \langle
v(t')v(t'')\rangle dt'dt''.
\end{equation}
Thus asymptotic power-law velocity correlations with an exponent
$\eta <1 $ are expected to imply asymptotically anomalous diffusion
of the form
\begin{equation}
\langle x^2 (t) \rangle  \sim t^\alpha \label{alpha}
\end{equation}
with
\begin{equation}
\alpha =2-\eta .
\end{equation}
This relation simply results from the two time integrations.

It is interesting to compare our model with other dynamical models
generating Tsallis statistics. Plastino
and Plastino\cite{plasti} and Tsallis and Bukmann\cite{bukmann}
study a generalized Fokker-Planck equation of the form
\begin{equation}
\frac{\partial P(x,t)}{\partial t}=- \frac{\partial}{\partial x}
(F(x)P(x,t))+ D \frac{\partial^2}{\partial x^2}P(x,t)^\nu
\label{buk}
\end{equation}
with a linear force $F(x)=k_1-k_2x$ and $\nu\not= 1$. Basically this model
means that the diffusion constant becomes dependent on the
probability density. The probability densities generated by
eq.~(\ref{buk}) are $q$-exponentials with the exponent
\begin{equation}
q=2-\nu.
\end{equation}
The model generates anomalous diffusion with $\alpha = 2/(3-q)$.
Assuming the validity of $\alpha = 2-\hat{\eta}$, i.e. the generation
of anomalous diffusion by slowly decaying velocity correlations
with exponent $\hat{\eta}$, one obtains
\begin{equation}
\hat{\eta} =\frac{4-2q}{3-q}.
\end{equation}
On the other hand, for the $\chi^2$-superstatistical Langevin
model one obtains by combining eq.~(\ref{eta}) and (\ref{qqq}) the
different relation
\begin{equation}
\eta =\frac{5-3q}{2q-2}.
\end{equation}
Interesting enough, there is a distinguished $q$-value where both models
yield the same answer:
\begin{equation}
q=1.453 \Rightarrow \hat{\eta} =\eta =0.707
\end{equation}
These values of $q$ and $\eta$
correspond to realistic, experimentally observed numbers,
for example in defect turbulence\cite{daniels}.

So far we mainly studied correlation functions with power law
behaviour. But in fact one can construct superstatistical
Langevin models that exhibit more complicated types of asymptotic
behaviour of the correlation functions. To see this we notice
that the asymptotic analysis of section 3 applies to correlation
functions as well, by formally defining
\begin{eqnarray}
\tilde{E}&:=&\frac{1}{2} \sigma^2 m|t-t'| \\
\tilde{f} (\beta) &=& \frac{1}{m\beta} f(\beta)
\end{eqnarray}
and writing
\begin{equation}
C(t-t')=\int_0^\infty \tilde{f} (\beta) e^{-\beta \tilde{E}}d\beta .
\end{equation}
To obtain statements on the symptotic decay rate of the
superstatistical correlation function, we may just use the same
techniques described in section 3 with the replacement $E\to
\tilde{E}$ and $f \to \tilde{f}$. In this way one can construct
models that have, for example, stretched exponential asymptotic
decays of correlations etc. (see also \cite{corr}). Asymptotic
means here that $|t-t'|$ is large as compared to the local
equilibrium relaxation time scale, but still smaller than the
superstatistical time scale $T$, such that the adiabatic
approximation is valid.

\section{Some Applications}

We end this paper by briefly mentioning some recent applications
of the superstatistics concept. Rizzo and
Rapisarda\cite{rapisarda,rap2} study experimental data of wind
velocities at Florence airport and find that
$\chi^2$-superstatistics does a good job. Jung and
Swinney\cite{jungswinney} study velocity differences in a
turbulent Taylor-Couette flow, which is well described by
lognormal superstatistics. They also find a simple scaling
relation between the superstatistical parameter $\beta$ and the
fluctuating energy dissipation $\epsilon$. Paczuski et
al.\cite{pac1} study data of solar flares on various time scales
and embedd this into a superstatistical model based on
$\chi^2$-superstatistics = Tsallis statistics. Human behaviour
when sending off print jobs might also stand in connection to
such a superstatistics\cite{pac2}. Bodenschatz et al.\cite{boden}
have detailed experimental data on the acceleration of a single test
particle in a turbulent flow, which is well described by
lognormal superstatistics, with a Reynolds number dependence as
derived in a superstatistical Lagrangian turbulence model studied
by Reynolds\cite{reynolds}. The statistics of cosmic rays is 
well described by $\chi^2$-superstatistics, with $n=3$ due to the
three spatial dimensions\cite{cosmic}. In mathematical finance
superstatistical techniques are well known
and come under the
heading `volatility fluctuations', see e.g.\cite{bouchard} for a
nice introduction and \cite{ausloos,econo} for some more recent
work. Possible applications also include granular media, which
could be described by different types of superstatistics,
depending on the boundary conditions\cite{vanzon}. The observed
generalized Tsallis statistics
of solar wind speed fluctuations\cite{burlaga} is a further candidate
for a superstatistical model.
Chavanis\cite{chavanis} points out analogies between
superstatistics and the theory of violent relaxation for
collisionless stellar systems. Most superstatistical models
assume that the superstatistical time scale $T$ is very large, so
that a quasi-adiabatic approach is valid, but Luczka and
Zaborek\cite{luczka} have also studied a simple model of
dichotomous fluctuations of $\beta$ where everything can be
calculated for finite time scales $T$ as well.

\end{document}